\documentclass{article}
\usepackage{natbib}
\usepackage{amsmath, epsfig}
\newtheorem{theorem}{Theorem}
\newcommand{\mat}[1]{\textrm{{{\textbf{#1}}}}}
\newcommand{\bm}[1]{\mathbf{#1}}

\begin{document}
\title{Multiple Testing with Heterogeneous Multinomial Distributions}\
\author{Joshua Habiger\footnote{Department of Statistics, Oklahoma State University, email:jhabige@okstate.edu},\\
David Watts\footnote{Department of Statistics, Oklahoma State University, email: david.watts@okstate.edu}, and \\
Michael Anderson \footnote{Department of Plant and Soil Science, Oklahoma State University, email: michael.anderson@okstate.edu}}\maketitle 
\begin{abstract}
False discovery rate (FDR) procedures provide misleading inference when testing multiple null hypotheses with heterogeneous multinomial data.  For example, in the motivating study the goal is to identify species of bacteria near the roots of wheat plants (rhizobacteria) that are associated with productivity, but standard procedures discover the most abundant species even when the association is weak or negligible, and fail to discover strong associations when species are not abundant.  Consequently, a list of abundant species is produced by the multiple testing procedure even though the goal was to provide a list of producitivity-associated species.  This paper provides an FDR method based on a mixture of multinomial distributions and shows that it tends to discover more non-negligible effects and fewer negligible effects when the data are heterogeneous across tests.  The proposed method and competing methods are applied to the motivating data.  The new method identifies more species that are strongly associated with productivity and identifies fewer species that are weakly associated with productivity. \\
\end{abstract}

Keywords: False Discovery Rate; Heterogeneity; Multinomial; Multiple hypothesis testing; Rhizosphere

\section{Introduction}
Bacteria that inhabit the interface between the root and the soil (the rhizosphere) play a crucial role in the growth and development of plants. The living portion of the rhizosphere is very complex: it is estimated to contain billions of individual organisms \citep{Gan05} and tens of thousands of species of bacteria (rhizobacteria) per gram of soil \citep{Men13}. Some rhizobacteria  positively impact plant health by processing soil nutrients into plant available forms \citep{Ves03}, helping plants tolerate stresses \citep{Yan09}, protecting plants against disease causing organisms \citep{Bak13} and improving soil structure \citep{Ala00}.  Others have a negative impact on plant growth \citep{Sus82}.  The identification of these rhizobacteria will allow for 1) the evaluation of current agricultural practices, 2) a more rigorous definition of productive and healthy soil systems and 3) the development of technologies to manipulate the rhizosphere for greater agricultural productivity and sustainability.

The goal in \cite{AndHab12} was to identify/discover species of rhizobacteria in wheat whose abundance is associated with productivity.  The statistical problem, formally described in Section \ref{sec motivating data} of this manuscript, amounts to testing multiple null hypotheses simultaneously with multinomial data. Standard statistical methods, outlined in Section \ref{sec standard approaches}, compute a test statistic for each null hypothesis and apply a false discovery rate (FDR) multiple testing procedure to the collection of test statistics. %The result is a list of rejected null hypotheses, also called discoveries, which for this data (seemingly) identifies or discovers the rhizobacteria that are associated with productivity.
For example, the well-known BH procedure \citep{BenHoc95} or an adaptive BH procedure \citep{Sto04, LiaNet12} can be applied to the $p$-values or the local FDR procedure, introduced in \cite{EfrTib01} and formalized in \cite{SunCai07}, could be applied to the $Z$-scores for the tests.

In Sections \ref{subsection significant} and \ref{sec data analysis} we see that these standard approaches provide misleading inference.  In particular, some rhizobacteria that appear to be strongly associated with productivity are not discovered while other rhizobacteria that are only negligibly associated with productivity are discovered.   A more careful inspection reveals that standard methods will, in general, tend to discover the most abundant rhizobacteria, even if they are only negligibly associated with productivity.  Consequently, some abundant rhizobacteria are mislabeled as productivity-associated rhizobacteria and many productivity-associated bacteria are not discovered because they are not abundant.  This confounding inference can have a significant detrimental impact on efforts to understand how the rhizosphere interacts with the environment.  A similar phenomenon was observed in \cite{SunMcL12}, where the goal was to identify high schools whose test scores were associated with socio-economic status, but standard approaches tended to identify the largest schools instead.

It turns out that standard methods fail because the sample sizes for the multinomial data are heterogeneous across tests.  Section \ref{sec clFDR} provides a multiple testing procedure for multinomial data which incorporates this heterogeneity via conditional local FDR (clFDR) statistics.  The statistics are referred to as ``conditional'' to emphasize the fact that analysis is based on a multinomial model for each test, which arises by conditioning on the (random) sample sizes.  The result is a procedure that depends functionally upon the sample sizes across tests that, as we will show, controls the FDR.  Section \ref{sec thresholding} shows analytically that the clFDR procedure provides more scientifically relevant inference and shows that the improvement can be characterized as a ``thresholding effect''.  That is, the rejection region for a test based on the clFDR statistic is decreased when the sample size is large and increased when the sample size is small.  Consequently, the probability of discovering a negligible association is decreased and the probability of discovering a strong association is increased. Section \ref{sec data analysis} applies the new method to the motivating data and verifies that it performs as anticipated.

\section{Motivating data and objective} \label{sec motivating data}
The basic objective is to identify rhizobacteria that are \textit{significantly} associated with wheat shoot biomass, which is a measure of wheat productivity.  For the moment we use the term ``significantly'' somewhat freely.  Clarification is provided in Section \ref{subsection significant}.

The motivating data are depicted in Table \ref{data}.  For a detailed account of the experiment see \cite{AndHab12}.  Here, $y_{nm}$ represents the number of DNA copies, or abundance, of the $m$th rhizobacterial species among the $n$th group of wheat plants for $n = 1, 2, ..., N$ and $m = 1, 2, ..., M$.  For this data $N = 5$ and $M = 778$.  The average shoot biomass among wheat plants in the $n$th group is denoted $x_n$ and the vector of biomass measurements is denoted by $\mat{x} = (x_1, x_2, ..., x_N)^T$.   Denote the vector of abundance measurements for species $m$ by $\mat{y}_m = (y_{1m}, y_{2m}, ..., y_{Nm})^T$ and the random vector by $\mat{Y}_m$.  Denote the $M\times N$ random matrix by $\mat{Y} = (\mat{Y}_1^T, \mat{Y}_2^T, ..., \mat{Y}_M^T)^T$. Denote the random sample size (row total in Table \ref{data}) for $\mat{Y}_m$ by $N_m = \sum_{n = 1}^NY_{nm}$ and a realization of $N_m$ by $n_m$.
\begin{table}
\caption{Depiction of the data in \cite{AndHab12}.  Shoot biomass $x_n$ in grams for groups $n$ = 1, 2, ..., 5 was 0.86, 1.34, 1.81, 2.37, and 3.00, respectively. Row totals are in the last column.}
\label{data}
\begin{center}
\begin{tabular}{ccccccc|c} \hline
Species $m$ && $y_{1m}$ & $y_{2m}$ & $y_{3m}$ & $y_{4m}$ & $y_{5m}$ & Total ($n_m$)\\ \hline
%\cline{1-1} \cline{3-8} \cline{3-8} \cline{1-1}
1 && 0 & 1 & 1 & 0 & 5 & 7\\
2 && 9 & 2 & 0 & 0 & 3 & 14\\
$\vdots$&&$\vdots$&$\vdots$&$\vdots$&$\vdots$&$\vdots$ & $\vdots$\\
778 && 16 & 10 & 29 & 18 & 13 & 81 \\
\hline
\end{tabular}
\end{center}
\end{table}

While more complex models could be considered, here we consider a log-linear model for each species for simplicity and to facilitate parameter estimation and mixture modeling later.  Specifically, assume $Y_{nm}$ has a Poisson distribution with mean $\mu_{nm}$ and assume that $log(\mu_{nm}) = \alpha_{m} + \beta_{m}x_n$, where each $\alpha_m$ and $\beta_m$ are regression parameters taking values in $\Re$. Further assume $Y_{1m}, Y_{2m}, ..., Y_{Nm}$ are independent for each $m$.  Observe that if $\beta_m = 0$ then $\mu_{nm} = \alpha_m$ for each $x_n$, and hence the abundance of species $m$ is not associated with productivity. If $\beta_m$ is positive/negative then $\mu_{nm}$ is increasing/decreasing in $x_n$ (note that $x_1<x_2<...<x_N$).  Thus, productivity-associated bacteria can be identified by testing the null hypothesis $H_m:\beta_m = 0$ for each of the $M = 778$ bacteria.   The unobservable state of a null hypothesis is denoted by $\theta_m = I(\beta_m\neq 0)$, where $I(\cdot)$ is the indicator function.  The decision to reject or retain $H_m$ using $\mat{Y}$ is denoted $\delta_m(\mat{Y}) = I(H_m \mbox{ rejected})$.

A $p$-value or $Z$-score for $H_m$ can be based on sufficient statistic $T_m = \mat{x}^T\mat{Y}_m$.  To avoid estimation of the nuisance parameter $\alpha_{m}$, tests based on this log-linear model often utilize the conditional distribution of $\mat{Y}_m|N_m=n_m$, which has multinomial probability mass function
\begin{equation}\label{pmf}
p(\mat{y}_m|n_m; \beta_m) = \frac{n_m!}{\prod_{n=1}^Ny_{nm}!}\prod_{n=1}^N p_{n}(\beta_m)^{y_{nm}} \mbox{ where }p_{n}(\beta_m) = \frac{\exp\{\beta_{m}x_n\}}{\sum_{n=1}^N\exp\{\beta_{m}x_n\}}.
\end{equation}
For details and additional motivation see \cite{McCNel89}. Denote the multinomial probability vector by $\mat{p}(\beta_m) = [p_1(\beta_m), p_2(\beta_m), ..., p_N(\beta_m)]^T$. Now, observe that \\$E[T_m|n_m; \beta_m] = n_m\mat{x}^T\mat{p}(\beta_m)$ where $E[T_m|n_m;\beta_m]$ denotes the expectation of $T_m$ taken with respect to the probability mass function in \eqref{pmf}, and 
$Var(T_m|n_m;\beta_m) = n_m\mat{x}^T\bm{\Sigma}(\beta_m)\mat{x}$ where $\bm{\Sigma}(\beta_m) =  [diag(\mat{p}(\beta_m)) - \mat{p}(\beta_m)\mat{p}(\beta_m)^T]$.  Thus, under $H_m:\beta_m = 0$ a $Z$-score can be computed
$$Z_m = \frac{T_m - n_m\mat{x}^T\mat{p}(0)}{\sqrt{n_m\mat{x}^T\bm{\Sigma}(0)\mat{x}}}.$$
A $p$-value for $Z_m$ can be computed $P_m = 2[1-F_0(|Z_m|)]$,
where $F_0(z)$ is the cumulative distribution function of $Z_m$ under Model \eqref{pmf} when $\beta_m = 0$.  Note that $F_0$ can be approximated with a standard normal distribution function or the exact distribution of $F_0$ can be simulated by sampling from $p(\mat{y}_m|n_m; \beta_m)$ with $\beta_m = 0$.

Typically a multiple decision function $\bm{\delta}(\mat{Y}) = [\delta_1(\mat{Y}), \delta_2(\mat{Y}), ..., \delta_M(\mat{Y})]^T$ which controls the FDR at some fixed level $\alpha$ is employed in this ``large $M$'' setting.  To define the FDR, denote the number of discoveries by $R(\mat{Y}) = \sum_{m=1}^M\delta_m(\mat{Y})$ and denote the number of false discoveries by $V(\bm{\theta},\mat{Y}) = \sum_{m=1}^M \delta_m(\mat{Y})(1-\theta_m)$, where $\bm{\theta} = (\theta_1, \theta_2, ..., \theta_M)^T$. Then
$$FDR = E\left[\frac{V(\bm{\theta},\mat{Y})}{R(\mat{Y})}\Big{|}R(\mat{Y})>0\right]\Pr(R(\mat{Y})>0)$$
where the expectation is taken over $\mat{Y}$.  See \cite{BenHoc95} or \cite{Sto03} for alternative definitions and discussions.

\section{Standard FDR procedures} \label{sec standard approaches}
Standard FDR procedures are defined in terms of the $Z$-scores or $p$-values.  For example, the well-known \cite{BenHoc95} procedure ranks the $p$-values $P_{(1)}\leq P_{(2)}\leq ... \leq P_{(M)}$ and rejects the $k$ null hypotheses with the smallest $p$-values, where $k = \max\{m:P_{(m)}\leq \alpha m /M\}$.  If $P_{(m)} >\alpha m/M$ for each $m$ then no null hypotheses are rejected.  Adaptive $p$-value procedures \citep{Sto04, LiaNet12} for FDR control operate in a similar fashion, but incorporate an estimate of the proportion of true null hypotheses.  See \cite{BlaRoq09} for a more comprehensive list and comparison of $p$-value procedures for FDR control.

Local FDR procedures based on $Z$-scores typically utilize a random mixture model, perhaps first considered within the context of the FDR in \cite{EfrTib01, GenWas02,Sto03}. For additional references see \cite{Efr10}.  For example, assume $Z_1, Z_2, ..., Z_M$ are independent and identically distributed with mixture density $f(z)$ defined by
$$f(z) = \pi_0 f_0(z) + (1-\pi_0) f_1(z),$$
where $\pi_0\in (0,1)$ is the mixing proportion and $f_0$ and $f_1$ are the densities when $H_m$ is true and false, respectively.  If the state of $H_m$ is viewed as random then $\pi_0$ is the prior probability that $H_m$ is true, i.e. $\pi_0  = \Pr(\theta_m = 0)$.  The local FDR under mixture density $f$ is defined by
%\begin{equation}\label{lFDR fxn}
$$lFDR(z;f) = \frac{\pi_0f_0(z)}{f(z)}$$
%\end{equation}
and the local FDR statistic for $Z_m$ is defined $lFDR_m = lFDR(Z_m;f)$.

The lFDR procedure in \cite{SunCai07} is operationally implemented as follows.  First, rank the lFDR statistics $lFDR_{(1)}\leq lFDR_{(2)}\leq ... \leq lFDR_{(M)}$.  If $\sum_{i = 1}^m lFDR_{(i)}>m\alpha$ for each $m$ then no null hypotheses are rejected.  Otherwise, the $k$ null hypotheses corresponding to $lFDR_{(1)}, lFDR_{(2)}, ..., lFDR_{(k)}$ are rejected, where
%\begin{equation}
$$\label{k}
k = \max\left\{m:\sum_{i = 1}^m lFDR_{(i)}\leq m\alpha\right\}.
$$
%\end{equation}
The lFDR procedure can also be written $\delta_m(\mat{Z}) = I(lFDR_m\leq \lambda)$, where
\begin{equation}
\lambda = \left\{\begin{array}{l c}
             0 & \mbox{ if } $k = 0$\\
             lFDR_{(k)}& \mbox{ otherwise.}
          \end{array}\right.
\end{equation}
While this thresholding notation seems redundant, it will be useful for studying lFDR procedures later.  \cite{SunCai07} showed that local FDR procedures dominate FDR methods based on $p$-value statistics in that among all procedures with asymptotic $FDR\leq \alpha$, they have the smallest missed discovery rate, defined $$MDR = \frac{E[S(\bm{\theta},\mat{Y})]}{E[M - R(\mat{Y})]}$$ where $S(\bm{\theta},\mat{Y}) = \sum_{m=1}^M\theta_m[1-\delta_m(\mat{Y})]$ is the number of erroneously retained null hypotheses and $M - R(\mat{Y})$ is the number of retained null hypotheses.  Hence we focus on lFDR procedures for the remainder of this manuscript.

In practice $f$ is not known but can be consistently estimated, thereby rendering the resulting lFDR procedure adaptive.  %\cite{SunCai07} showed that if $f$ is consistently estimated in the above procedure, then the resulting adaptive procedure has FDR that is less than or equal to $\alpha$ and has the smallest $MDR$ among all procedures providing FDR control, asymptotically.
To illustrate, assume $f$ is a mixture of 3, 4, or 5 normal densities with one of the component densities being a standard normal (null) density.  Maximum likelihood estimates (found using the mixtools package in R), the AIC and BIC for each model, and the number of discoveries made by the adaptive lFDR procedure are summarized in Table \ref{model summaries}.

\begin{table}
\caption{Parameter estimates, the AIC and BIC, and the number of discoveries $R(\mat{Y})$ when implementing the adaptive $lFDR$ procedure at $\alpha = 0.05$ are summarized when using a 3, 4, and 5 component normal mixture density, with one component density being the standard normal density.}
\label{model summaries}
\begin{center}
\begin{tabular}{lccc}\hline
 $\hat\pi_0, (\hat\pi_1,\hat\mu_1,\hat\sigma_1^2)$, $(\hat\pi_2,\hat\mu_2,\hat\sigma_2^2)$, $\ldots$ & AIC & BIC & $R(\mat{Y})$  \\  \hline
0.68, (0.22,-1.5,3.3$^2$), (0.10,1.9,0.9$^2$) &   3139.6 & 3167.6 & 85  \\
0.21, (0.01,-9.0,5.7$^2$), (0.11, -1.3, 2.6$^2$), (0.67,0.1,1.6$^2$)  & 3082.2 & 3124.1 & 175 \\
0.36, (0.2,-6.8,5.5$^2$), (0.53,-0.3,1.8$^2$), (0.07,1.1,0.5$^2$), (0.01, 2.7,0.3$^2$) & 3085.0 & 3141.0 & 125 \\ \hline
\end{tabular}
\end{center}
\end{table}

\section{Non-negligible effects}\label{subsection significant}
So which model should we use?  Observe that the 4 and 5 component mixture models could be deemed preferable over the 3 component model because they have lower AIC and BIC values and lead to more discoveries.  However, each of these models have a non-null component density that is very near the null standard normal density. For example, one of the densities in the 4 component model has mean $0.1$ and variance $1.6^2$ while one component density has mean -0.3 and variance $1.8^2$ for the 5 component model.   If the objective of the analysis is to discover \textit{non-negligible} effects, we may be tempted to reconsider these two models because a discovery could mean that the $Z$-score was generated from a density that is only negligibly different from the null density.

\begin{figure}
\begin{center}
\includegraphics[width = 5in]{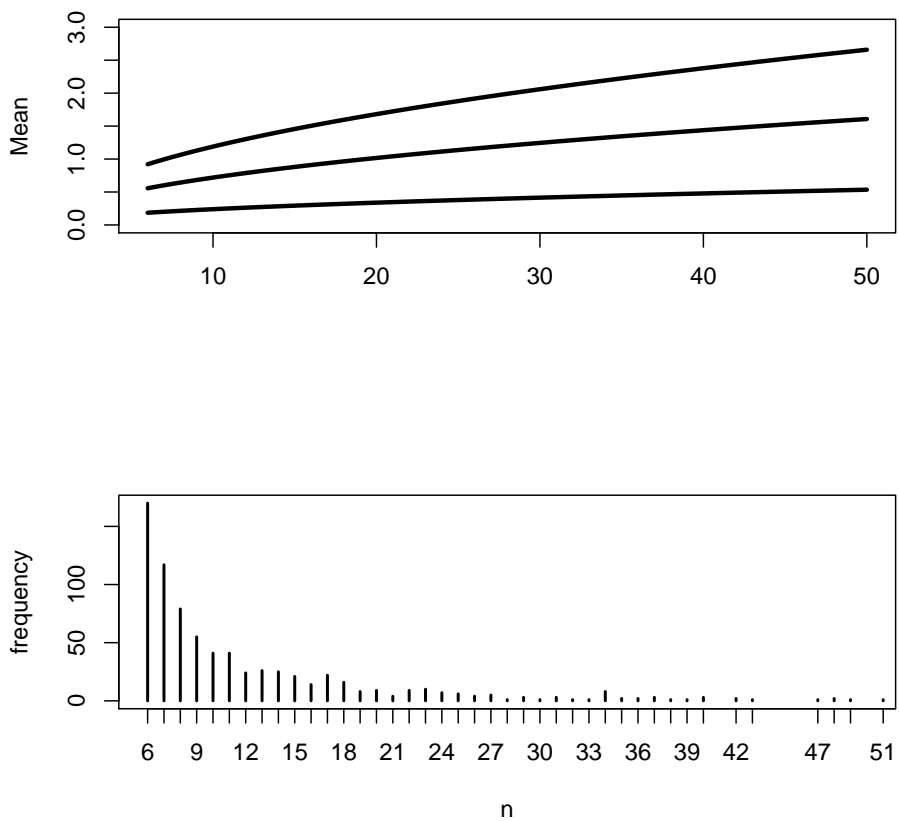}
\caption{Top: Plot of $\mu(n,\gamma)$ vs. $n$ for $\gamma = 0.1, 0.3, 0.5$ from bottom to top, respectively.  Bottom: Bar chart of $n_m$s.  The $n_m$s greater than 51 (27 of them)  are not depicted.}
\label{mean vs. n}
\end{center}
\end{figure}

However, a more careful inspection reveals that a species with a $Z$-score from a ``near-null'' component density might be more scientifically meaningful than $Z$-scores from a component density that is farther from the null and vice versa.    To see this, for $\beta_m = \gamma >0$ and $N_m = n$, denote the conditional mean of $Z_m$ under model \eqref{pmf} by
\begin{equation}\label{mean}
\mu(n,\gamma)\equiv E[Z_m|N_m = n;\beta_m = \gamma] = \frac{\sqrt{n}\mat{x}^T[\mat{p}(\gamma)-\mat{p}(0)]}{
\sqrt{\mat{x}^T\bm{\Sigma}(0)\mat{x}}}
\end{equation}
and the conditional variance by
\begin{equation}\label{variance}
\sigma^2(\gamma) \equiv Var(Z_m|N_m = n; \beta_m = \gamma) = \frac{\mat{x}^T\bm{\Sigma}(\gamma)\mat{x}}{\mat{x}^T\bm{\Sigma}(0)\mat{x}}.
\end{equation}
Note that $\mat{p}(0.1)^T = (0.18, 0.19, 0.20, 0.21, 0.22)$ deviates from the null probabilities $\mat{p}(0) = \mat{0.2}$ by a small, perhaps negligible, amount.  When $\gamma = 0.3$ or $\gamma = 0.5$, which we refer to as moderate effects and large effects, respectively, probabilities are farther from $\mat{0.2}$.  For example, $\mat{p}(0.5)^T = (0.11, 0.14, 0.18, 0.24, 0.33)$.  Figure \ref{mean vs. n} plots $\mu(n,\gamma)$ vs. $n$ for $\gamma = 0.1, 0.3, 0.5$ and displays the distribution of the $n_m$s for the data in Table \ref{data}.  Observe that even for moderate effects, $Z_m$ has mean between 0 and 1 when $n_m$ is less than or equal to 10  and $n_m\leq 10$ for the majority (59\%) of tests.  On the other hand, when $n$ is large $\mu(n,\gamma)$ can be large even for negligible effects, and $n$ can be as large as 911. For example, $\mu(911,0.1) = 2.28$.  For moderate effects, $\mu(911,0.3) = 6.86$.  Hence, we may anticipate a procedure based on $Z$-scores alone to reject $H_m$ when $n_m$ is large even when $\beta_m$ is only negligibly different from 0.  Likewise, $H_m$ will likely be retained even when $\beta_m$ is significantly different (in a practical sense) from 0 when $n_m$ is small.  For this particular data set, this means that species which are merely the most abundant (large $n_m$) will tend to be identified as productivity-associated even if the association is weak or negligible, while species that are strongly correlated with productivity may not be discovered because they are not abundant.

The fundamental issue is that any procedure based on $Z$-scores (or $p$-values) alone can only be used to detect large $|\mu(n_m,\beta_m)|$, which may or may not indicate that the parameter of interest $\beta_m$ is negligibly different from 0.  Indeed, for \textit{any} $\beta_m \neq 0$, $|\mu(n_m,\beta_m|$ is monotonically increasing in $n_m$.  Hence for large enough $n_m$ we should anticipate $|Z_m|$ to be large even if $\beta_m$ is only negligibly different from 0.  Next we develop an FDR procedure which allows for more direct inference on $\beta_m$.

\section{A conditional local FDR procedure}\label{sec clFDR}
The proposed procedure is based on a mixture of multinomial distributions.  We first define the procedure assuming parameters in the model are known.  Parameter estimation follows.
\subsection{The procedure}
Assume that $\mat{Y}_m|N_m = n_m$ is generated from a mixture of multinomial probability mass functions.  Specifically, let $\bm{\gamma} = (\gamma_0, \gamma_1, ..., \gamma_K)^T$ be the collection of $K+1$ distinct possible values for $\beta_m$ in $\Re$ with $K\geq 1$ and $\gamma_0 = 0$.  Denote mixing proportions by $\bm{\pi} = (\pi_0, \pi_1, ..., \pi_K)^T$ satisfying $\sum_{k=0}^K \pi_k = 1$ and $0<\pi_k<1$ for each $k$.  Define the mixture probability mass function
\begin{equation}
\label{pmf mixture}
p(\mat{y}_m|n_m;\bm{\gamma},\bm{\pi}) = \sum_{k = 0}^K \pi_kp(\mat{y}_m|n_m;\gamma_k) = \sum_{k=0}^K\pi_k \frac{n_m!}{\prod_{n=1}^Ny_{nm}!}\prod_{n=1}^N p_{n}(\gamma_k)^{y_{nm}}
\end{equation}
where $p_n(\gamma_k)$ is the multinomial probability from the log-linear model defined as in \eqref{pmf}. Denote $\theta_m = I(\beta_m \neq \gamma_0)$ so that the $m$th null hypothesis is $H_m:\theta_m = 0$ or equivalently $H_m:\beta_m = \gamma_0$.  Observe that $\pi_0$ is still the proportion of true null hypotheses or the prior probability that $H_m$ is true.

Given parameters $\bm{\pi},\bm{\gamma}$, define
%\begin{equation}\label{lfdr multinomial}
$$clFDR(\mat{y}_m,n_m;\bm{\pi},\bm{\gamma}) = \frac{\pi_0p(\mat{y}_m|n_m;\gamma_0)}{p(\mat{y}_m|n_m;\bm{\gamma},\bm{\pi})}
$$
%\end{equation}
and define the conditional lFDR statistic for $H_m$ by $clFDR_m = clFDR(\mat{Y}_m,N_m;\bm{\pi},\bm{\gamma}).$ To define the clFDR procedure, first rank the clFDR statistics $clFDR_{(1)}\leq clFDR_{(2)}\leq ... \leq clFDR_{(M)}$.  If $\sum_{i = 1}^m clFDR_{(i)}>m\alpha$ for each $m$, reject $k = 0$ null hypotheses.  Otherwise, reject the $k = \max\{m:\sum_{=1}^mclFDR_{(i)}\leq m\alpha\}$ null hypotheses corresponding to $clFDR_{(1)}, clFDR_{(2)}, ..., clFDR_{(k)}$.  Formally, the $m$th clFDR decision function is defined $\delta_m^{c}(\mat{Y}) = I(clFDR_m\leq \lambda)$, where
$$
\lambda = \left\{\begin{array}{l c}
             0 & \mbox{ if } $k = 0$\\
             clFDR_{(k)}& \mbox{ otherwise.}
          \end{array}\right.
$$
and denote the clFDR multiple decision function by $\bm{\delta}^c(\mat{Y}) = [\delta_1^c(\mat{Y}), \delta_2^c(\mat{Y}), ..., \delta_M^c(\mat{Y})]^T.$

Under the model in \eqref{pmf mixture} the clFDR procedure controls the FDR. A proof is formally presented below for completeness, but is essentially the same as the proof of Theorem 1 in \cite{SunCai09} with a few minor notational modifications.  The important point is that the proof carries over as long as $clFDR_m$ is the posterior probability that the null hypothesis is true, i.e. $clFDR_m = \Pr(\theta_m = 0|\mat{y}_m,n_m;\bm{\pi},\bm{\gamma})$. %For example, we can also compute the posterior probability $lFDR_m = \Pr(\theta_m = 0|\mat{y}_m;\bm{\pi},\bm{\gamma})$, and the proof still carries over.
\begin{theorem}
For $0<\alpha\leq 1$, $\bm{\delta}^c(\mat{Y})$ has $FDR\leq \alpha$ under the model in \eqref{pmf mixture}.
\end{theorem}
\textit{Proof:
Observe that given $\mat{Y}_m = \mat{y}_m$ and $N_m = n_m$, $clFDR_m = \Pr(\theta_m = 0|\mat{y}_m,n_m;\bm{\pi},\bm{\gamma})$ and that if $R(\mat{Y}) = k>0$ then $E[V(\bm{\theta},\mat{Y})|\mat{Y}, R(\mat{Y})=k] = \sum_{i = 1}^k clFDR_{(i)}\leq k\alpha$ by construction. Thus, by the law of iterated expectation we have
\begin{eqnarray*}
FDR &=&E\left[\sum_{k = 1}^M\frac{1}{k}E\left[V(\bm{\theta},\mat{Y})|\mat{Y}, R(\mat{Y}) = k\right]P(R(\mat{Y})=k)\right]\\
&\leq& E\left[\sum_{k = 1}^M \frac{\alpha k }{k}\Pr(R(\mat{Y}) = k)\right] \leq \alpha.
%&=& \sum_{k = 1}^{M}\left[\frac{1}{k}\sum_{i = 1}^k lFDR_{(i)}^c\right]\Pr(R(\mat{Y}) = k) \\
%&\leq&\sum_{k = 1}^M\alpha\Pr(R(\mat{Y}) = k) \leq \alpha
\end{eqnarray*}
}
%It is important to note that the proof is valid under model \eqref{pmf mixture}, which does not make any assumption regarding the dependence structure of $\mat{Y}_1, \mat{Y}_2, ..., \mat{Y}_M$.

As in the previous section, parameters must be estimated to implement the clFDR procedure. That is, the proposed adaptive clFDR procedure is the same as the clFDR procedure, except that it uses adaptive clFDR statistics, defined $\widehat{clFDR}_m = clFDR(\mat{y}_m, n_m;\hat{\bm{\pi}},\hat{\bm{\gamma}})$ for each $m$ where $\hat{\bm{\pi}}$ and $\hat{\bm{\gamma}}$ are estimates of $\bm{\pi}$ and $\bm{\gamma}$.  Maximum likelihood estimation is discussed next.  %Then, each adaptive clFDR decision function is defined $\delta_m^{ac}(\mat{Y}) = I(\widehat{lFDR}_m^c \leq \lambda)$ where $\lambda$ is computed as in \eqref{get c} using the adaptive clFDR statistics.   The EM algorithm for estimating parameters is outlined next.

\subsection{Maximum likelihood estimation}

Denote $\bm{n} = (n_1, n_2, ..., n_M)^T$ and assume that given $\mat{n}$, $\mat{Y}$ has joint probability mass function
$$p(\mat{y}_1, \mat{y}_2, ..., \mat{y}_M|\mat{n}; \bm{\gamma},\bm{\pi}) \equiv \prod_{m=1}^Mp(\mat{y}_m|n_m;\bm{\gamma},\bm{\pi}) = \prod_{m=1}^M\sum_{k = 0}^K\pi_k p(\mat{y}_m|n_m;\gamma_k).$$ Define log likelihood function
$$l(\bm{\gamma},\bm{\pi}) \equiv \sum_{m = 1}^M\log\left(\sum_{k=0}^K\pi_kp(\mat{y}_m|n_m;\gamma_k)\right).$$
To facilitate the EM algorithm \citep{Dem77}, let $\mat{z}$ be a $(K+1)\times M$ matrix with $z_{km}$ taking on value 1 if $\mat{y}_m$ has component pmf $p(\mat{y}_m|n_m;\gamma_k)$ and 0 otherwise.  Then the complete data log likelihood is
$$l_c(\bm{\gamma},\bm{\pi}) = \sum_{m = 1}^M\sum_{k = 0}^Kz_{km}\log (\pi_k p(\mat{y}_m|n_m;\gamma_k)).$$

Now, let $\bm{\gamma}^{old}$ and $\bm{\pi}^{old}$ denote current parameter estimates of $\bm{\gamma}$ and $\bm{\pi}$, respectively.  Then the E and M steps below yield new parameter estimates, denoted $\bm{\gamma}^{new}$ and $\bm{\pi}^{new}$.
\begin{itemize}
\item[E:] For $k = 0, 1, ..., K$ and $m = 1, 2, ..., M$ compute $$\hat z_{km} = \frac{\pi_k^{old} p(\mat{y}_m|n_m;\gamma_k^{old})}{\sum_{k = 0}^K\pi_k^{old}p(\mat{y}_m|n_m;\gamma_k^{old})}.$$
\item[M:] Find $\bm{\gamma}^{new}$ and $\bm{\pi}^{new}$, the values of $\bm{\gamma}$ and $\bm{\pi}$, respectively, that maximize
\begin{eqnarray}
Q(\bm{\gamma},\bm{\pi}) &\equiv& \sum_{m = 1}^M\sum_{k = 0}^K\hat z_{km}\log (\pi_k p(\mat{y}_m|n_m;\gamma_k)) \nonumber \\
 &=&\sum_{m = 1}^M\sum_{k = 0}^K\hat z_{km}\log (\pi_k) + \sum_{m = 1}^M\sum_{k = 0}^K\hat z_{km}\log p(\mat{y}_m|n_m;\gamma_k) \label{Q}
\end{eqnarray}
subject to constraint $\sum_{k = 0}^K\pi_k = 1$.
\end{itemize}
Maximizing the first quantity in \eqref{Q} via Lagrangian optimization gives $$\bm{\pi}^{new} = \frac{1}{M}\sum_{m = 1}^M \hat{\mat{z}}_{m},$$ where $\hat{\mat{z}}_m = (\hat{z}_{0m}, \hat{z}_{1m}, ..., \hat{z}_{Km})^T$.  To get $\bm{\gamma}^{new}$, note that the second quantity in \eqref{Q} is proportional to
\begin{equation}\label{optim fxn}
 \sum_{k=1}^K g_k(\gamma_k) \equiv \sum_{k = 1}^K\sum_{m = 1}^M\hat z_{km}\sum_{n = 1}^Ny_{nm}\left[\gamma_kx_n - \log\left(\sum_{n = 1}^N \exp\{\gamma_kx_n\}\right)\right]\\
\end{equation}
The expression in \eqref{optim fxn} can now be maximized using any standard optimization method, such as a Newton-Rhapson routine.  In fact it is important to note that a one-dimensional optimization routine can be applied $K$ times to maximize each $g_k(\gamma_k)$, i.e.  multi-dimensional optimization can be avoided. Now, the E and M steps can be repeated until, say,
$l(\bm{\gamma}^{new},\bm{\pi}^{new}) - l(\bm{\gamma}^{old},\bm{\pi}^{old})<\epsilon$, yielding maximum likelihood estimators denoted $\hat{\bm{\gamma}}$ and $\hat{\bm{\pi}}$.

\section{The thresholding effect} \label{sec thresholding}
This section demonstrates that conditional lFDR procedures tend to discover more non-negligible effects and fewer negligible effects than the usual (unconditional) lFDR procedures. The main result is first motivated using the data in Table \ref{data}.

To facilitate a simple and broadly applicable comparison of clFDR and lFDR procedures, we consider model \eqref{pmf mixture} with two components, i.e. $\beta_m = 0$ or $\beta_m = \gamma_1>0$.  We also utilize a normal approximation for each test statistic $T_m = \mat{x}^T\mat{Y}_m$.  Recall that for $\beta_m = \gamma_1 > 0$ and conditionally upon $N_m = n_m$, the $Z$-score for $T_m$ has mean $\mu(n_m,\gamma_1)$ and variance $\sigma^2(\gamma_1)$.  See expressions \eqref{mean} and \eqref{variance}, respectively.  Thus, by the central limit theorem and delta method, given $N_m = n_m$ and $\beta_m = \gamma_1$, $Z_m$ is asymptotically normal (as $n_m \rightarrow \infty$) with mean $\mu(n_m,\gamma_1)$ and variance $\sigma^2(\gamma_1)$. Here, $\theta_m = I(\beta_m = \gamma_1)$ and again denote mixing proportions by $\pi_0$ and $\pi_1$.  In this section we suppress $\bm{\gamma}$ and $\bm{\pi}$ in the notation when possible for brevity.

The clFDR and lFDR procedures for comparison are based on normal approximations for the conditional and marginal mixture densities for $Z$-scores above.  Specifically, denote the conditional mixture density by
\begin{equation}\label{conditional mixture}
f(z|n) = \pi_0\phi(z;0,1) + \pi_1\phi(z;\mu(n,\gamma_1), \sigma^2(\gamma_1)),
\end{equation}
where $\phi(\cdot;a,b)$ denotes a normal density function with mean $a$ and variance $b$. Define $$clFDR(z,n) = \frac{\pi_0\phi(z;0,1)}{f(z|n)}.$$
%The marginal mixture density corresponding to the Model in \eqref{conditional mixture}, note that $E[Z_m|\theta_m = 0] = 0$ and $Var(Z_m|\theta_m = 1)$.  Further, by the law of iterated expectation, $\mu_1(\gamma_1) \equiv E[Z_m|\theta_m=1] = E[\mu(N_m,\gamma_1)]$ and
%\begin{eqnarray*}
%\sigma_1^2(\gamma_1) \equiv Var(Z_m|\theta_m = 1)&=& E[Var(Z_m|N_m,\theta_m = 1)] + Var(E[Z_m|N_m,\theta_m = 1])\\
% &=& \sigma^2(\gamma_1) + Var(\mu(N_m,\gamma_1))
%\end{eqnarray*}
Assume that $N_1, N_2, ..., N_M\stackrel{i.i.d.}\sim p(n)$ and denote the support of $N_m$ by $\mathcal{N}$.  Then the marginal mixture density for \eqref{conditional mixture} is
\begin{equation}\label{marginal mixture}
f(z) = \pi_0\phi(z;0,1) + \pi_1\sum_{n\in\mathcal{N}}\phi(z;\mu(n,\gamma_1), \sigma^2(\gamma_1))p(n)
\end{equation}
and the (marginal) local FDR is
$$lFDR(z) = \frac{\pi_0 \phi(z;0,1)}{f(z)}.$$
Observe that $f(z)$ and $lFDR(z)$ necessarily depend upon $p(n)$.  In this section we use the empirical probability mass function depicted in Figure \ref{mean vs. n} for $p(n)$ in all computations, unless otherwise specified.

\begin{figure}
\begin{center}
\includegraphics[width = 2in]{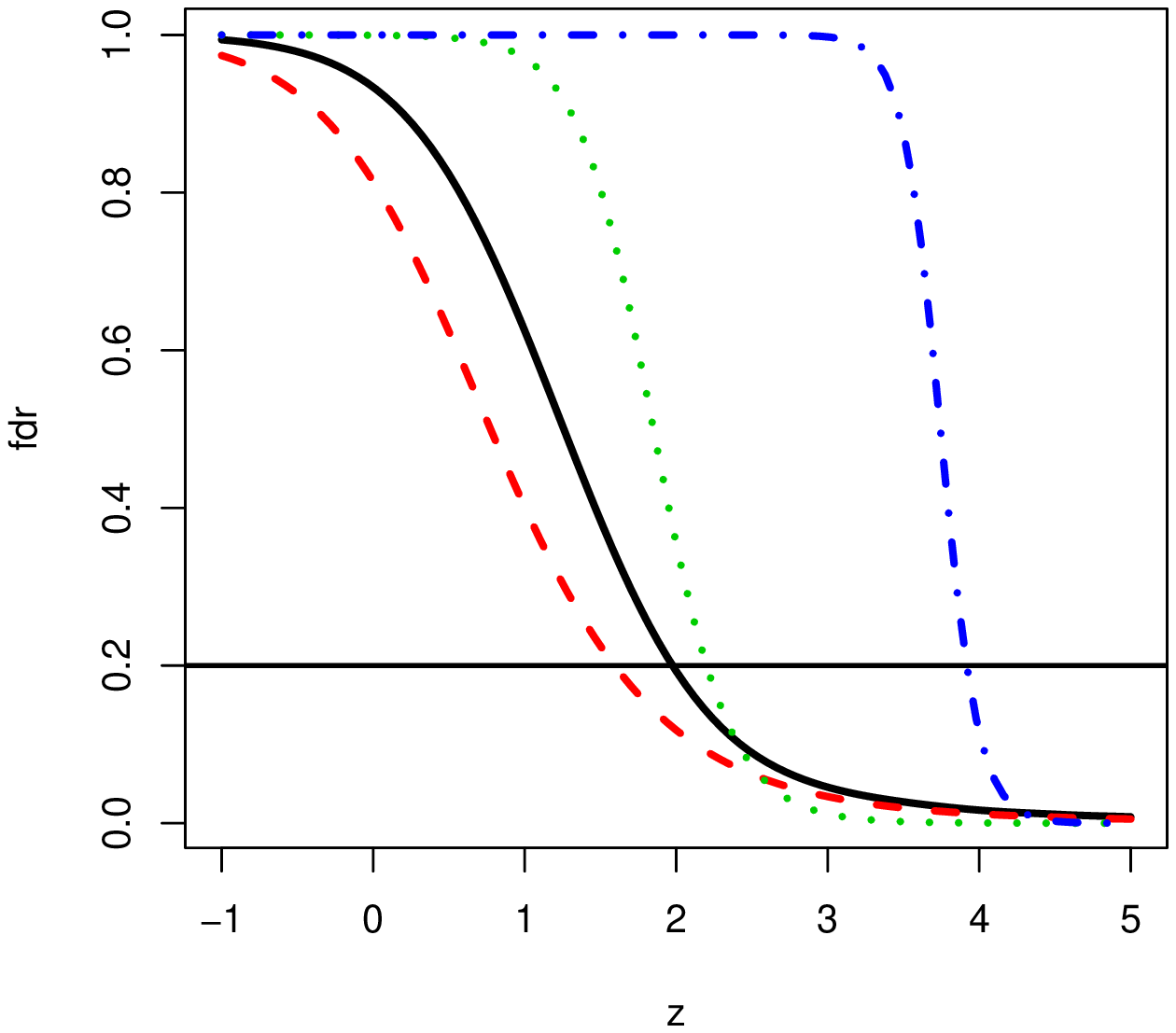}
\includegraphics[width = 2in]{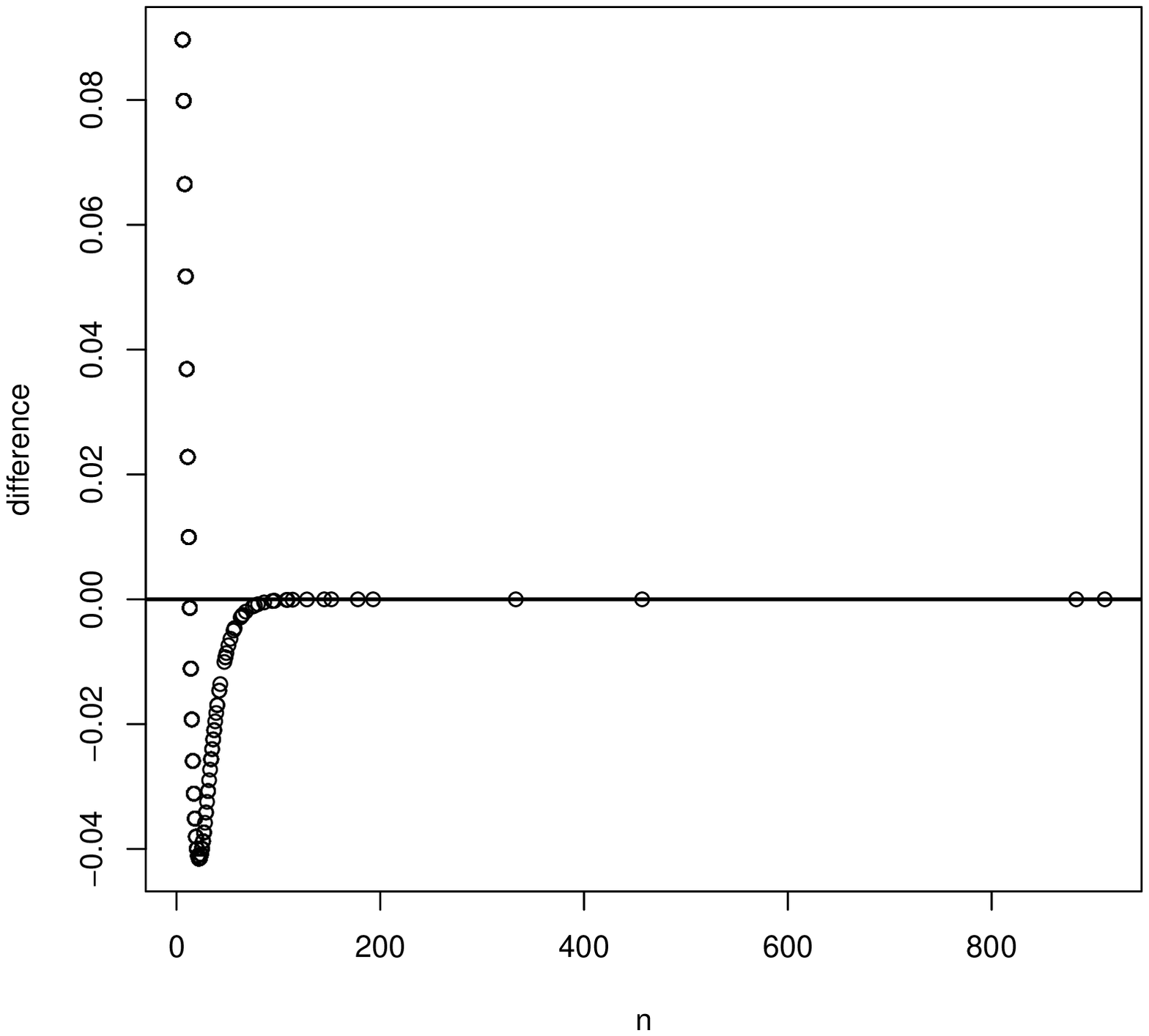}
\caption{Left: $clFDR(z,n)$ vs. $z$ for $n = 5$ (- - -), $n = 25$ ($\cdot$ $\cdot$ $\cdot$), and $n = 100$ (- $\cdot$ -) and $lFDR(z)$(---) vs. $z$ for $\gamma_1 = 1$ and $\pi_0 = 0.5$.  Right: difference in correct rejection probabilities (clFDR - lFDR) vs. $n_m$ for $n_m$ in Table \ref{data}. }
\label{Oracle plot}
\end{center}
\end{figure}

The left panel in Figure \ref{Oracle plot} plots $lFDR(z)$ and $clFDR(z,n)$ vs. $z$ for $\pi_0 = 0.5$, $\gamma_1 = 1$, $n = 5, 25, 100$.  Recall that lFDR and clFDR procedures reject $H_m$ if $clFDR(z_m,n_m)\leq \lambda$ or if $lFDR(z_m)\leq \lambda$ for some cutoff $\lambda$.  For sake of illustration, we take $\lambda = 0.2$ in what follows.  Observe in Figure \ref{Oracle plot} that $lFDR(z)\leq 0.2$ if $z>1.98$ regardless of $n$.  See the Appendix for exact expressions and discussion regarding the approximate equivalence of these rejection regions.   The important point is that even though the rejection regions do not depend on $n_m$, the probability of correctly rejecting $H_m$ (the power) is large when $n_m$ is large and small when $n_m$ is small due to the fact that $\mu(n_m,\gamma_1)$ increasing in $n_m$.  For example, if $n_m = 5$ then under model (\ref{conditional mixture}) $Z_m$ has a normal distribution with mean $\mu(5,1) = 1.59$ and variance $\sigma^2(1) = 0.89^2$ when $H_m$ is false.  Thus the power is $1-\Phi(1.98; 1.59, 0.89^2) = 0.33$, where $\Phi(\cdot;a,b)$ is the normal cumulative distribution function with mean $a$ and variance $b$.  When $n_m = 25$, $\mu(25,1)$ is 3.55 and the power is 0.96.   When $n_m = 100$, the power is 1.00.  This same phenomenon occurs for any $\gamma_1>0$, including values of $\gamma_1$ that are only negligibly different from 0.

Procedures based on the conditional lFDR dampen this effect by decreasing the threshold for rejection when $n_m$ is small and increasing it when $n_m$ is large.  For example, when $n_m = 5$, $clFDR(z_m,n_m) \leq 0.2$ when $z_m\geq1.6$.  That is, the $Z$-score threshold for rejection based on the clFDR is smaller than the $Z$-score threshold based on the lFDR.  Consequently, the power of the test based on the clFDR is larger in this small $n_m$ setting; it is 0.5 rather than 0.41.  Now suppose that $n_m = 25$.  Then $clFDR(z_m,n_m)\leq 0.2$ if $z_m\geq 2.2$.  Because the $Z$-score threshold for the clFDR procedure is now larger than the threshold for the lFDR procedure, the power for the clFDR procedure is smaller: it is 0.94 rather than 0.98.  In summary, conditioning on the $N_m$s in the local FDR analysis ensures that $Z$-score thresholds for rejection increase in $n_m$.  Consequently, the probability of discovering a negligible effect due to $n_m$ being large is decreased and the probability of discovering a non-negligible effect when $n_m$ is small is increased.  See the right panel of Figure \ref{Oracle plot}.

Theorem \ref{thm threshold} states that the above phenomenon, where thresholds for rejection increase as $n_m$ increases, occurs for large enough $n_m$ as desired.  See the Appendix for the proof and a discussion regarding the implied approximation of the rejection region $[clFDR(z,n)\leq \lambda]$ with $[z\geq a(n)]$.
\begin{theorem}\label{thm threshold}
Let $a(n)$ denote the smallest solution to $clFDR(z,n) = \lambda$.  Under model \eqref{conditional mixture}, $a(n)$ is increasing in $n$ whenever
\begin{equation} \label{increasing threshold}
\mu(n, \gamma_1)^2> 2 \log\left(\sigma(\gamma_1)\frac{\pi_0(1-\lambda)}{(1-\pi_0)\lambda}\right).
\end{equation}
\end{theorem}

\begin{figure}
\begin{center}
\includegraphics[width = 5in]{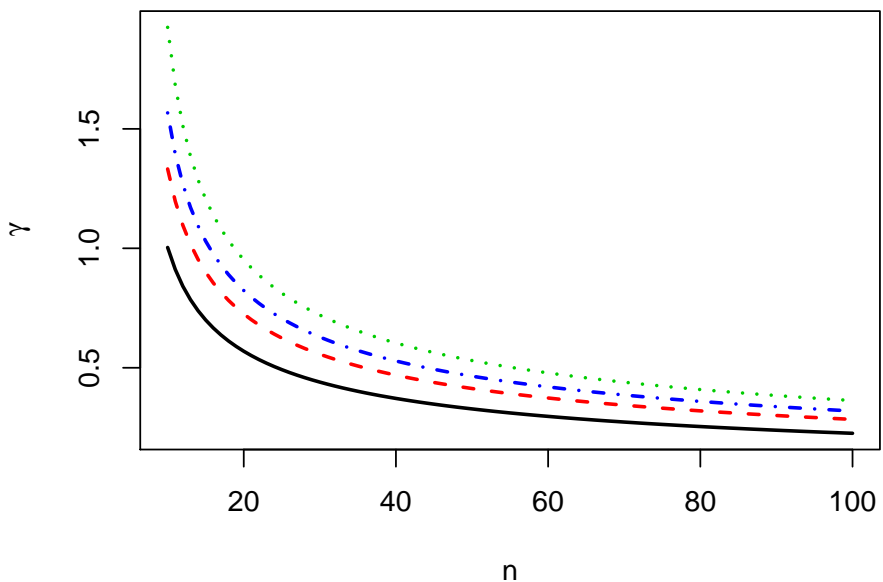}
\caption{Depiction of $(n,\gamma_1)$ values satisfying equation \eqref{increasing threshold}. Lines represent values of $n$ and $\gamma_1$ where the inequality in \eqref{increasing threshold} is an equality for $(\lambda,\pi_0) = (0.1, 0.8), (0.2, 0.8),  (0.1,0.5), (0.2,0.5)$ from top to bottom.  The inequality is satisfied whenever $(n,\gamma_1)$ is to the right of the line. }
\label{thm2 plot}
\end{center}
\end{figure}

Lets consider some specific settings.  First recall that $\lambda = clFDR_{(k)}$ satisfies \\$\frac{1}{k}\sum_{i = 1}^k clFDR_{(i)}\leq \alpha$. That is the average clFDR among the $k$ rejected null hypothesis is near $\alpha$.  Hence, the threshold $\lambda = clFDR_{(k)}$ should be larger than $\alpha$, especially if the clFDR statistics have a skewed right distribution. Hence, we consider $\lambda = 0.1$ and $\lambda = 0.2$.  We also consider $\pi_0 = 0.5$ and $0.8$ in our discussion.  Each line in Figure \ref{thm2 plot} represents values of $n$ and $\gamma_1$  where the inequality in \eqref{increasing threshold} is an equality.  The inequality is satisfied whenever $(n,\gamma_1)$ is to the right of a line.  For example, when $\lambda = 0.2$ and $\pi_0 = 0.5$, we see that the inequality is satisfied for every $n\geq 10$ if $\gamma_1 = 1$.  If $\gamma_1 = 0.5$ instead of $1$, then the inequality is satisfied for $n$ greater than or equal to 25. When $\lambda = 0.1, \pi_0 = 0.5, $ and $\gamma_1 = 0.5$ so that tests are generally less powerful, the inequality is satisfied for $n\geq 35$.  The fact that the thresholds don't begin to increase until n is moderate in low power settings (smaller $\gamma$ and $\lambda$) is to be anticipated as the power of the test is still relatively small when $n$ is moderate.  Hence, the probability of discovering a negligible effect is still small for moderate $n$.

\section{Application} \label{sec data analysis}
Here we analyze the data in Table \ref{data} using the adaptive clFDR procedure in Section \ref{sec clFDR} and compare the results to the adaptive lFDR procedure results for the $Z$-scores in Section \ref{sec standard approaches}.  Both procedures are applied at $\alpha = 0.05$.  For the adaptive clFDR procedure, we consider a 3 and 4 component mixture model and use convergence criterion $l(\bm{\gamma}^{new},\bm{\pi}^{new}) - l(\bm{\gamma}^{old},\bm{\pi}^{old})<10^{-8}$.  The EM algorithm failed to converge after 1000 iterations for a 5 component mixture model.  Parameter summaries and the number of discoveries are in Table \ref{analysis table}. R code for implementing the EM algorithm is available upon request.
\begin{table}
\caption{Parameter estimates, AIC, BIC and the number of discoveries for Models 1 and 2. }
\label{analysis table}
\begin{center}
\begin{tabular}{cccccccc}
\hline
Model & $\hat\pi_0$ & $(\hat\pi_1,\hat\gamma_1)$ & $(\hat\pi_2, \hat\gamma_2)$ & $(\hat\pi_3, \hat\gamma_3)$ & AIC & BIC & R(\mat{Y}) \\ \hline
1 & 0.69 & (0.16, -1.13) & (0.15,0.78) & NA & 12221.9& 12240.5 & 99\\
2 & 0.69 & (0.03, -2.68)&(0.13, -1.03) & (0.15, 0.79) & 12145.6 & 12173.5 & 97\\ \hline \\
\end{tabular}
\end{center}
\end{table}

\begin{figure}
\begin{center}
\includegraphics[width=5in]{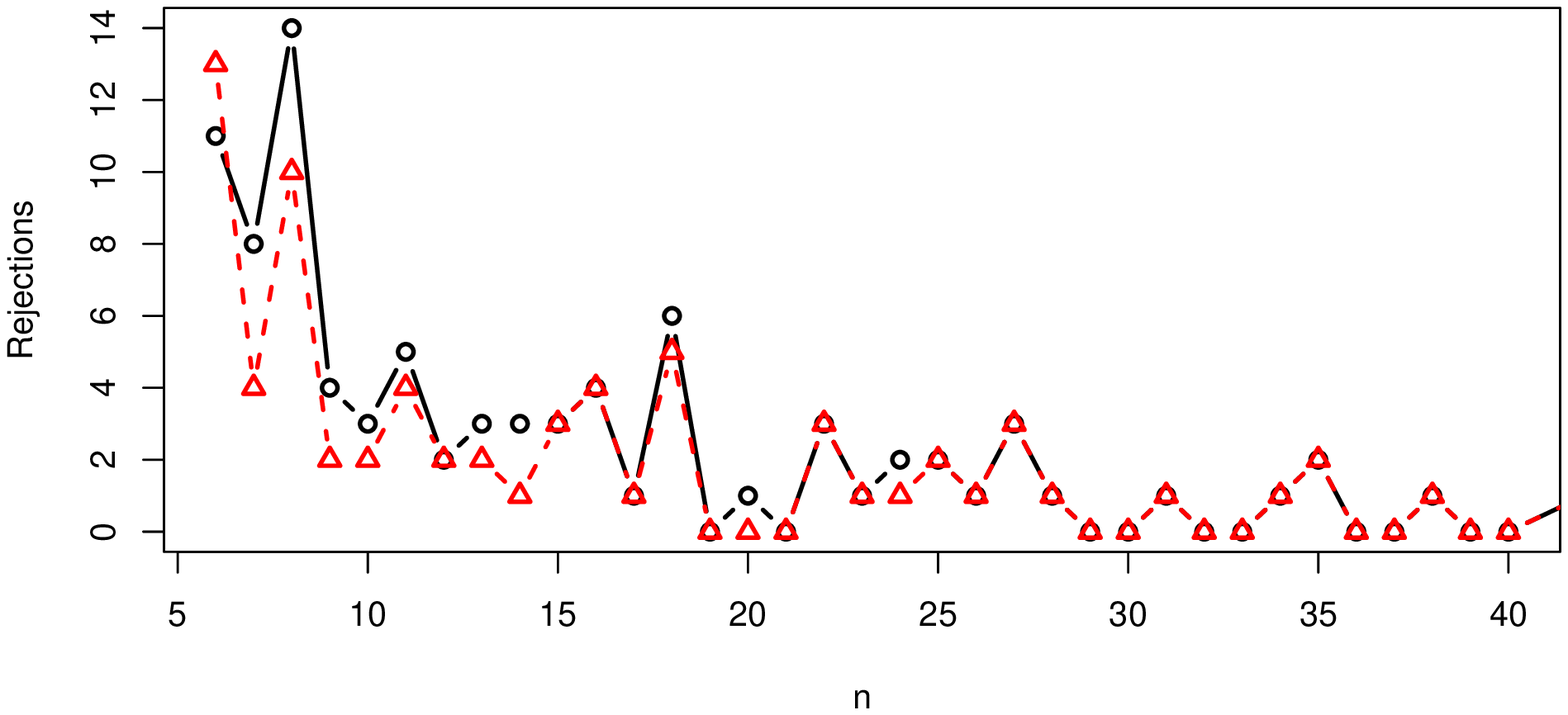}
\caption{The number of rejections vs. n for the conditional lFDR procedure (o) and for the lFDR procedure ($\triangle$). }
\label{rejections}
\end{center}
\end{figure}

Observe that all non-null component probability mass functions signify practical significance regardless of the number of components considered.   That is, $\hat{\gamma}_k$ is more than arbitrarily different from 0.  For example, $\mat{p}(0.78) = [0.08, 0.11, 0.16, 0.25, 0.40]$ certainly deviates from $\mat{p}(0) = \bm{0.2}$ by an amount that would signify a non-negligible association with productivity.

A more careful inspection of the results confirms that the clFDR procedure tends to discover more non-negligible effects and fewer negligible effects.  Figure \ref{rejections} plots the number of discoveries vs. $n$ for the 3 component normal mixture model in Table \ref{model summaries} and for the 3 component multinomial mixture model in Table \ref{analysis table}.  Observe that the clFDR procedure makes more discoveries when $n_m$ is small, as Theorem \ref{thm threshold} suggests.  Also, when $n_m\geq 40$ (not plotted), the clFDR procedure makes two fewer discoveries.  These latter discrepancies occur for $n_m = 193$ and $n_m = 911$.

In general, the characteristics of the data when discrepancies exist are as anticipated.  For example, $\mat{Y}_m^T/n_m = (5, 7, 0, 1, 1)/11 = (0.56, 0.64, 0, 0.09, 0.09)$ is certainly more than negligibly different from $\mat{p}(0) = \bm{0.2}$, but was declared significant only by the clFDR procedure.  Here $n_m$ is small.   On the other hand, $\mat{Y}_m^T/n_m = (134, 117, 252, 231, 177)/911 = (0.15, 0.13, 0.28, 0.25,0.19)$ deviates from $\mat{p}(0)$ less severely but $n_m$ is large.  Consequently, it was discovered by the lFDR procedure but not by the clFDR procedure.

%It should be noted that the naive procedure did made 2 additional rejections when $n_m$ was small.  In each setting the data were $\mat{Y}_m^T/n_m = (0, 0, 0, 3, 3)$.  Here, the $Z$-score was 2.63 and the sample cell probabilities, though they deviate from $\bm{0.2}$ by a signficant amount, they also are not close to $\mat{p}(\gamma_1)$ or $p(\gamma_2)$.

\section*{Appendix}
First we verify that for $\sigma^2(\gamma_1) \neq 1$, the rejection region $\{z: clFDR(z,n)\leq \lambda\}$ can be written $\{z: a(n)\leq z \leq b(n)\}$ and then demonstrate that it is well approximated by $\{z: z\geq a(n)\}$ under model \eqref{conditional mixture}.  Expressions for $a(n)$ and $b(n)$ are also derived.  For brevity, denote $\mu(n,\gamma_1)$ by $\mu(n)$ and $\sigma^2(\gamma_1)$ by $\sigma^2$.

Observe that setting $clFDR(z,n) = \lambda$, plugging expressions for the normal densities in $clFDR(z,n)$, and rearranging we get
$$z^2(\sigma^2 -1) + 2z\mu(n) - 2\sigma^2 log(\sigma k) - \mu(n)^2 = 0,$$
where $k = \frac{\pi_0(1-\lambda)}{(1-\pi_0)\lambda}$.  The solutions of this quadratic ($a(n)$ and $b(n)$) are
$$\frac{\mu(n) \pm \sqrt{\mu(n)^2\sigma^2 - 2(1-\sigma^2)\sigma^2\log(\sigma k)}}{1-\sigma^2}.$$

To verify that  $\{z: clFDR(z,n)\leq \lambda\} = \{z: a(n)\leq z \leq b(n)\}$, first observe that taking the derivative of $clFDR(z,n)$ with respect to $z$ gives
$$clFDR'(z,n) = \left[\frac{\pi_0(1-\pi_0)\phi(z;0,1)\phi(z;\mu(n),\sigma^2)}{f(z|n)^2}\right]\left[\frac{z(1-\sigma^2) -\mu(n)}{\sigma^2}\right]$$
which is negative for $z<\mu(n)/(1-\sigma^2)$ and positive for $z>\mu(n)/(1-\sigma^2)$.  Hence, $clFDR(z,n)$ is ``U'' shaped and hence $clFDR(z,n)\leq \lambda$ if and only if $a(n)\leq z\leq b(n)$.

It should be noted that if $\sigma^2 = 1$ then $clFDR'(z,n)\leq 0$ for $z\in \Re$ and the rejection region is of the form $[z>a]$ for some $a$. It can be verified, however, that $\sigma^2<1$ whenever $\gamma\neq 0$, thereby resulting in the above rejection region.  This rejection rule is seemingly untractable because we fail to reject $H_m$ if $Z_m>b(n)$ even though $Z_m>b(n)$ is stronger evidence against $H_m$ than $a(n)\leq Z_m\leq b(n)$.  However, the event that $Z>b(n)$ occurs with very small probability in practice and hence has little impact on the clFDR procedure.  For example, $[b(n) - \mu(n)]/\sigma$ was at least 4.6 for all combinations of $\gamma = 0.5, 1, 2; \lambda = 0.05,0.1,0.2; n = 5, 100, 1000;$ and $\pi_0 = 0.1, 0.5, 0.9$ and is typically greater than 50. That is, $b(n)$ was at least 4.6 standard deviations above the mean when $H_m$ was false, and hence the probability of such a $Z$-score is extremely small. The curious reader is referred to \cite{Cao13} for a more detailed investigation.

\noindent\textbf{Proof of Theorem \ref{thm threshold}}: The derivative of $a(n)$ with respect to $\mu(n)$ is
$$ \frac{1}{1-\sigma^2}\left[1 - \frac{\mu(n)\sigma^2}{\sqrt{\mu(n)^2\sigma^2 - 2(1-\sigma^2)\sigma^2log(\sigma k)}}\right]$$ which is positive when
$$\frac{\mu(n)\sigma^2}{\sqrt{\mu(n)^2\sigma^2 - 2(1-\sigma^2)\sigma^2log(\sigma k)}}<1.$$  Some algebra gives that this inequality is satisfied if and only if
$\mu(n)^2>2\log(\sigma k)$.  Plugging in $\pi_0(1-\lambda)/[(1-\pi_0)\lambda]$ for $k$ gives the desired inequality.

\bibliographystyle{chicago}
\bibliography{localFDR}

\end{document}